\documentclass[preprint,authoryear,12pt]{article}

\usepackage{amssymb}
\usepackage{amsmath, epsfig}
\usepackage{units, color,natbib}

\newcommand{\la}{\label}
\newcommand{\be}{\begin{equation}}
\newcommand{\en}{\end{equation}}

\renewcommand{\vec}[1]{\boldsymbol{#1}}
\newcommand{\ii}{\textrm{i}}
\newcommand{\ee}{\textrm{e}}

\def \div{\mbox{div\hskip 1pt}}
\def \Div{\mbox{Div\hskip 1pt}}
\def \tr{\mbox{tr\hskip 1pt}}
\def \grad{\mbox{grad\hskip 1pt}}
\def \Grad{\mbox{Grad\hskip 1pt}}
\begin{document}

\title{On  deforming a sector of a circular cylindrical\\  tube into an intact tube:\\ existence, uniqueness, and stability}

\author{M. Destrade$^{a,b}$, J.G. Murphy$^c$, R.W. Ogden$^{b,d,e}$ \\[12pt]
   $^a$School of Mathematics, Statistics and Applied Mathematics,  \\ 
   National University of Ireland Galway, \\
   University Road, Galway, Ireland \\[12pt]
   $^b$School of Electrical, Electronic, and Mechanical Engineering, \\
University College Dublin, \\ 
Belfield, Dublin 4, Ireland\\[12pt]
  $^c$Department of Mechanical Engineering, \\
 Dublin City University, \\
 Glasnevin, Dublin 9, Ireland\\[12pt]
  $^d$ School of Engineering,\\
    University of Aberdeen,  \\
    King's College, Aberdeen AB24 3UE, United Kingdom\\[12pt]
 $^e$School of Mathematics and Statistics,  \\
 University of Glasgow, \\
 University Gardens, Glasgow G12 8QW, Scotland, UK}
 \date{}

\maketitle

\newpage

\begin{abstract}

Within the context of finite deformation elasticity theory the problem of deforming an open sector of a thick-walled circular cylindrical tube into a complete circular cylindrical tube is analyzed.  
The analysis provides a means of estimating the radial and circumferential residual stress present in an intact tube, which is a problem of particular concern in dealing with the mechanical response of arteries.  
The initial sector is assumed to be unstressed and the stress distribution resulting from the closure of the sector is then calculated in the absence of loads on the cylindrical surfaces. 
Conditions on the form of the elastic strain-energy function required for existence and uniqueness of the deformed configuration are then examined. 
Finally, stability of the resulting finite deformation is analyzed using the theory of incremental deformations superimposed on the finite deformation, implemented in terms of the Stroh formulation. 
The main results are that convexity of the strain energy as a function of a certain deformation variable ensures existence and uniqueness of the residually-stressed intact tube, and that bifurcation can occur in the closing of thick, widely opened sectors, depending on the values of geometrical and physical parameters. 
The results are illustrated for particular choices of these parameters, based on data available in the biomechanics literature.

\end{abstract}




\section{Introduction}


Residual stresses have a very important role in the mechanical functioning of arteries and the existence of residual stresses is well documented, as in, for example, \cite{Chuo86}, \cite{Vais83} and \cite{Fung93}; see also the review by \cite{Hump02}.
They are demonstrated by the simple device of cutting radially a short ring of an artery, the result of the cut being the springing open of the ring into a sector, thereby releasing some residual stresses; in general, some residual stress remains, however, as shown by \cite{Voss93} and \cite{Gree97}; see also the review by \cite{Rach03}.
A crude estimate of the circumferential residual stresses is obtained by measuring the resulting angle of the sector into which the ring deforms, although it has to be said that reliable quantitative data remain elusive.
This so-called \emph{opening angle method} has been analyzed in some detail by several authors, including, for example, \cite{Delf97}, \cite{Zidi99}, \cite{Rach99}, \cite{Holz00}, \cite{Ogde00}, \cite{MaSa02}, \cite{Ogde03}, \cite{Ragh04} and \cite{Olss06}. 
The typical approach is to assume that the (unloaded) opened sector is free of stress and circular cylindrical and then to construct a deformation that brings the sector into an intact tube and to calculate the resulting radial and circumferential (residual) stress distributions.  
This has been done for a single layer and for two-layered tubes.  
In general this method does not account for any stress or deformation in the axial direction.  
However, in reality, not only does the ring open into a sector, but the length of the arterial segment tends to change and there are also residual stresses in the axial direction, although quantitative information about axial residual stresses is distinctly lacking.  A recent 3D analysis by \cite{Holz10} is the first attempt to calculate the combination of radial, circumferential and axial residual stresses. The purpose of the present paper is to provide an analysis of the opening angle method, with particular reference to the questions of existence and uniqueness of the residually-stressed configuration and its stability.

In Section \ref{sec2} we review the description of the geometry of the deformation which first takes the sector into an intact cylinder, allowing for a possible length change, and then allows inflation under internal pressure combined with axial load while maintaining circular cylindrical symmetry.
The corresponding stress components and equilibrium equations are then summarized in respect of an isotropic elastic material whose properties are described in terms of a strain-energy function.
The existence and uniqueness of the resulting configuration is then examined in Section \ref{sec3}, under the assumption that the strain energy is convex as a function of a suitably chosen deformation variable.  
The Mooney-Rivlin and the Fung strain energies obey this assumption, and we use them, as well as some experimental data on arteries, to illustrate the analysis. 
Section \ref{sec4} is devoted to an analysis of the stability of the residually stressed tube (in the absence of lateral loads) on the basis of the theory of small incremental deformations superimposed on a finite deformation, and for this purpose an appropriate version of the Stroh formulation is adopted.  
For simplicity of illustration, attention is restricted to prismatic incremental deformations. 
We then treat numerically the case of tubes made of solids with the Mooney-Rivlin strain energy, and compare the predictions to some simple experiments we performed with silicone rubber.


\section{Problem formulation\label{sec2}}



\subsection{Geometry of the deformation}


We consider an annular sector of a circular cylindrical tube with geometry defined by
\begin{equation} \label{region}
A \le R \le B,  \quad
- (2\pi- \alpha)/2 \le \Theta \le + (2\pi - \alpha)/ 2,
\quad
0\le Z\le L,
\end{equation}
where $A$ and $B$ are the radii of the inner and outer faces, respectively, of the sector, $L$ is its length, and  $\left( R, \Theta, Z \right)$ denote the cylindrical coordinates of a point in the material, with corresponding orthonormal basis vectors ($\vec{E}_R$, $\vec{E}_\Theta$, $\vec{E}_Z$). 
Here $\alpha\in (0,2\pi)$ is the so-called \emph{opening angle}. 
The sector is assumed to be unstressed in this reference configuration, which we denote by $\mathcal{B}_0$.

\begin{figure}[!t]
\center
\epsfig{figure=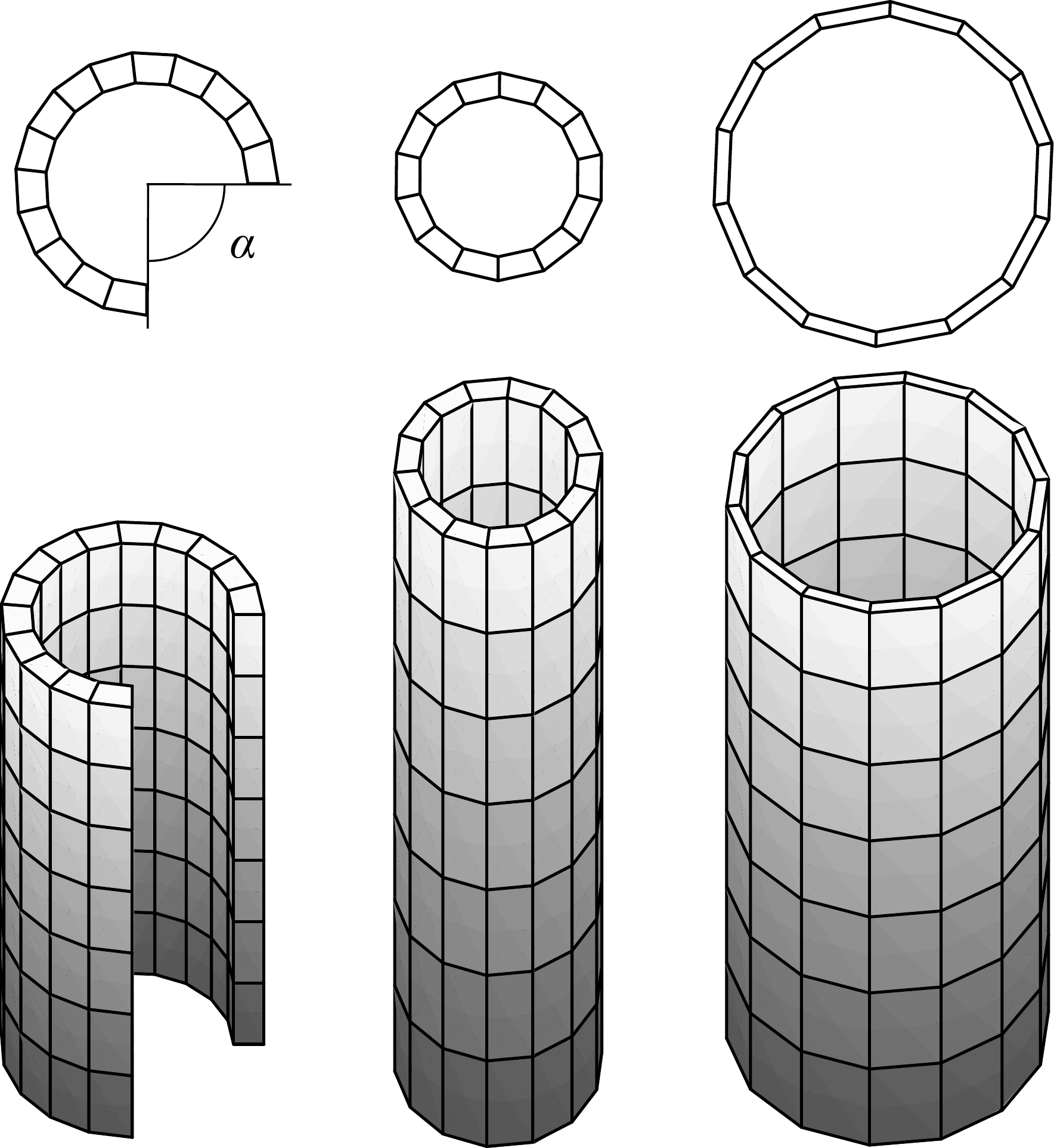, width=.8\textwidth}
 \caption{Depiction of the three configurations: (a) unstressed open sector (configuration $\mathcal{B}_0$), with the opening angle $\alpha$ shown as $\pi/2$; (b) residually stressed and stretched tube (configuration $\mathcal{B}_r$); (c) tube in (b) subject to internal pressure $P$
(configuration $\mathcal{B}_l$).}
 \label{figure1}
\end{figure}

The sector is then deformed into an intact (circular cylindrical) tube so that the plane faces originally at $\Theta =\pm (\pi -\alpha  / 2)$ are joined perfectly together and there is an accompanying uniform axial stretch $\lambda_z = l/L >1$, where $l$ is the deformed length of the tube.
We refer to this new configuration as the \emph{residually-stressed} configuration, which we denote by $\mathcal{B}_r$.  
In this configuration there is no traction on the curved surfaces of the tube, but in general axial loads are required to maintain the deformation since it turns out that the axial stress depends on the radius $r$ and cannot therefore be set to zero pointwise on the ends of the tube.  
Finally, the tube is subjected to a uniform internal pressure of magnitude $P$ per unit deformed area and is maintained in a new circular cylindrical configuration, which we refer to as the \emph{loaded  configuration}, denoted $\mathcal{B}_l$; the three separate configurations are depicted in Figure \ref{figure1}.

Let $\left(r, \theta, z \right)$ denote cylindrical coordinates in either the residually-stressed configuration $\mathcal{B}_r$ or the pressurized configuration $\mathcal{B}_l$, with  corresponding orthonormal basis vectors ($\vec{e}_r$, $\vec{e}_\theta$, $\vec{e}_z$).  
The deformation may be described by the equations
\be
r= r\left(R\right), \quad
\theta = k \Theta, \quad
z=  \lambda_z  Z,
\en
from which we may calculate the deformation gradient tensor $\vec{F}=\Grad\vec{x}$ as
\be \label{F}
\vec{F} = r'  \vec{e}_r \otimes \vec{E}_R +  (k r/R)   \vec{e}_\theta \otimes \vec{E}_\Theta + \lambda_z \vec{e}_z \otimes \vec{E}_Z,
\en
where the prime denotes differentiation and
\be
k \equiv \dfrac{2 \pi}{2\pi - \alpha}, \quad k > 1,
\end{equation}
is \emph{a measure of the  opening angle}.  The left Cauchy--Green tensor is then calculated as
\begin{equation}
\vec{B}=\vec{FF}^T={r'}^2\vec{e}_r\otimes\vec{e}_r+(kr/R)^2\vec{e}_\theta\otimes\vec{e}_\theta+\lambda_z^2\vec{e}_z\otimes\vec{e}_z,
\end{equation}
and it follows immediately that the principal axes of $\vec{B}$ are $\vec{e}_r,\vec{e}_\theta$, and $\vec{e}_z$. 
The corresponding principal stretches are
\be \label{lambdas1}
\lambda_1 = r', \quad
\lambda_2 = \dfrac{k r}{R} , \quad
\lambda_3 = \lambda_z.
\en

Arterial wall tissue is generally considered to be essentially \emph{incompressible}, so that the constraint $\det \vec{F} = 1$ is enforced.  For the present deformation this constraint yields
\begin{equation} \label{r^2}
r^2 = \dfrac{R^2 - A^2}{k \lambda_z} + a^2,
\end{equation}
where $a \equiv r(A)$ is the radius of the inner surface of the tube in configuration $\mathcal{B}_l$.
Similarly, we define $b \equiv r(B)$ for the outer surface.
It then follows from from \eqref{lambdas1} that the  principal stretches become
\be \label{lambdas2}
\lambda_1 = \dfrac{R}{k \lambda_z r}, \quad
\lambda_2 = \dfrac{k r}{R} , \quad
\lambda_3 = \lambda_z.
\en

Notice from \eqref{r^2} that if $a^2 k \lambda_z = A^2$ then the principal stretches are constants.
In other words, the deformation is \emph{homogeneous}, of the form
\be \label{hom}
r = R/ \sqrt{k \lambda_z}, \quad
\theta = k \Theta, \quad z = \lambda_z Z,
\en
with principal stretches $\lambda_1=1/\sqrt{k \lambda_z}$, $\lambda_2= \sqrt{k/\lambda_z}$, $\lambda_3=\lambda_z$.


\subsection{Stress components and equilibrium}


Henceforth we restrict the analysis to homogeneous, incompressible, iso\-tro\-pic hyperelastic solids. 
The strain-energy function is then a symmetric function of the principal stretches, which we write as $W(\lambda_1,\lambda_2,\lambda_3)$, subject to the constraint $\lambda_1\lambda_2\lambda_3=1$. For such materials the Cauchy stress tensor $\vec{\sigma}$ is coaxial with $\vec{B}$ and for the considered deformation the principal Cauchy stress components are
\begin{equation} \label{T}
\sigma_{rr} = -q +\lambda_1 \dfrac{\partial W}{\partial \lambda_1}, \quad
\sigma_{\theta \theta } = -q + \lambda_2 \dfrac{\partial W}{\partial \lambda_2}, \quad
\sigma_{zz} = -q + \lambda_3 \dfrac{\partial W}{\partial \lambda_3},
\end{equation}
where $q$ is a Lagrange multiplier introduced by the constraint of incompressibility, and $\lambda_1$, $\lambda_2$, $\lambda_3$ are given by \eqref{lambdas2}.

Because the stretches are independent of $\theta$, and $z$ the $\theta$ and $z$ components of the equilibrium equation ensure that $q$ is also independent of $\theta$ and $z$.  
The only non-trivial equation remaining is the radial equation
\be \label{equil}
\frac{\mathrm{d}}{\mathrm{d}r} \sigma_{rr}  +\frac{1}{r} (\sigma_{rr} - \sigma_{\theta \theta}) = 0.
\en
With this equation we associate the boundary conditions
\be \label{BC}
\sigma_{rr} (a) = -P, \quad \sigma_{rr} (b) = 0,
\en
where $P$ is the internal pressure in configuration $\mathcal{B}_l$ and the outer boundary is free of traction.

At this point it is convenient to introduce the notations
\be
x_a \equiv k\lambda_z \dfrac{a^2}{A^2}, \quad
x \equiv k\lambda_z \dfrac{r^2}{R^2}, \quad
x_b \equiv k\lambda_z \dfrac{b^2}{B^2},\quad \varepsilon \equiv \dfrac{B^2}{A^2} - 1 > 0 ,
\en
$\varepsilon$ being a measure of the thickness of the initial annular sector.
Then it follows easily that
\begin{equation} \label{gamma}
x_b = \dfrac{\varepsilon +  x_a}{\varepsilon +1}.
\end{equation}

For any fixed value of the axial stretch $\lambda_z$ we may now regard $W$ as a function of $x$ through the stretches, which are given in terms of $x$ by
\begin{equation}
\lambda_1=\frac{1}{\sqrt{k\lambda_z x}},\quad \lambda_2=\sqrt{\frac{kx}{\lambda_z}},
\end{equation}
together with $\lambda_3=\lambda_z$, and we write $\widehat{W}(x)=W(1/\sqrt{k\lambda_z x},\sqrt{kx/\lambda_z},\lambda_z)$.

From \eqref{T} it then follows that
\begin{equation}  \label{TT}
\sigma_{\theta \theta} - \sigma_{rr} = \lambda_2 \dfrac{\partial W}{\partial \lambda_2} -\lambda_1\dfrac{\partial W}{\partial \lambda_1}   =
 2 x  \widehat{W}'\left(x\right).
\end{equation}

Noting that
\be \label{change}
r\frac{\mathrm{d}x}{\mathrm{d}r} = 2 x(1- x),
\en
we may integrate equation \eqref{equil} and apply the boundary conditions \eqref{BC},  and thus conclude that
\begin{equation} \label{P}
P =
\int _{x_a}^{x_b} \dfrac{ \widehat{W}'\left(x\right)}{1 -  x} \mathrm{d}x.
\end{equation}
This equation, together with \eqref{gamma}, enables the location of the deformed inner radius to be determined for given values of the pressure $P$, the thickness (as measured by $\varepsilon$), the opening angle (as measured via $k$), and the axial stretch $\lambda_z$.
For completeness we compute the corresponding stress distribution, given by
\begin{equation} \label{Tdistrib}
\sigma_{rr} = \int_{x_b}^{x} \dfrac{ \widehat{W}'\left(s\right)}{1 - s} \mathrm{d}s,
\quad
\sigma_{\theta \theta } =\sigma_{rr}  + 2 x  \widehat{W}'\left(x\right),
\quad
\sigma_{zz} = \sigma_{rr} + \lambda_3 \dfrac{\partial W}{\partial \lambda_3} - \lambda_1\dfrac{\partial W}{\partial \lambda_1}.
\end{equation}

In the special case of the homogeneous deformation \eqref{hom}, we have $x\equiv 1$ and we find from \eqref{TT} that
\begin{equation}
\sigma_{\theta \theta} - \sigma_{rr}   =
 2   \widehat{W}'(1) =  \text{constant},
\end{equation}
so that we can integrate the equation of equilibrium \eqref{equil} to give
\be \label{Trr_hom}
\sigma_{rr} =
 2 \widehat{W}'(1) \ln \left(r/b \right),
\end{equation}
where the boundary condition \eqref{BC}$_2$ has been used.
The boundary condition \eqref{BC}$_1$ then yields
\be \label{Phom}
P=   \widehat{W}'(1) \ln \left(b/a\right)
=  \widehat{W}'(1) \ln \left( B/A \right).
\end{equation}
This is the pressure that needs to be applied inside the tube in order to produce a homogeneous deformation through the tube wall.
We note that in this case, both $\sigma_{rr}$, given by \eqref{Trr_hom}, and $\sigma_{\theta\theta}$, given by
\be
\sigma_{\theta \theta}  = P\left[1+\ln(r/b)\right]/\ln(b/a),
\en
have logarithmic variations. Hence, in the case of homogeneous deformation, the stress is a slowly varying function of the radial coordinate.

It is clear that in general $\sigma_{zz}$ depends on $r$.  
In the case of homogeneous deformations this follows immediately from \eqref{Trr_hom} and \eqref{Tdistrib}$_3$.  Thus, we emphasize that the pointwise end condition $\sigma_{zz}=0$ cannot be adopted and the residually stressed configuration $\mathcal{B}_r$ is not load free.  
In some situations it may be possible to enforce the condition that the resultant axial load vanishes, but here we merely consider that the axial stretch is fixed, so that the axial load has to be adjusted accordingly to accommodate this.


\subsection{Convexity of the strain-energy function}


Because we are considering an incompressible isotropic elastic material, the strain-energy function can also be written as $W= \widetilde{W}\left(I_{1} ,I_{2} \right)$, where $I_1$, $I_2$ are the first and second principal invariants of the Cauchy--Green deformation tensor $\vec{B}$.  Expressed in terms of the principal stretches these are
\begin{equation}  \label{I}
I_1 = \lambda_1^2 + \lambda_2^2 + \lambda_3^2, \quad
I_2 = \lambda_1^2 \lambda_2^2 + \lambda_1^2 \lambda_3^2 + \lambda_2^2 \lambda_3^2.
\end{equation}
From \eqref{lambdas2} we then have explicitly
\begin{align}
& I_1
= \left(\dfrac{R}{k \lambda_z r} \right)^2 + \left(\dfrac{k r}{R} \right)^2 + \lambda_z^2
= \dfrac{1}{k\lambda_z x} + \dfrac{k x}{\lambda_z} + \lambda_z^2,
\notag \\
& I_2 = \dfrac{1}{\lambda_z^2} + \left(\dfrac{R}{k r} \right)^2 + \left(\dfrac{k \lambda_z r}{R} \right)^2
= \dfrac{\lambda_z}{k x} + k \lambda_z x + \lambda_z^{-2}.
\end{align}

The function $ \widehat{W}'\left( x\right)$, occurring in the integrand of \eqref{P}, can therefore be written as
\begin{equation} \label{W'}
 \widehat{W}'\left(x \right) =
\dfrac{k^2 x^2 - 1}{k \lambda_z x^2} \left(\dfrac{\partial  \widetilde W}{\partial I_1} +  \lambda_z^2 \dfrac{\partial  \widetilde W}{\partial I_2} \right).
\end{equation}
Assuming that the standard empirical inequalities $\partial  \widetilde W / \partial I_1 >0$, $\partial  \widetilde W /  \partial I_2 \geq 0$ hold, an immediate conclusion is that independently of the form of $\widetilde{W}\left(I_{1} ,I_{2} \right)$
\begin{equation}  \label{xm}
 \widehat{W}'\left(x\right) = 0 \quad \Leftrightarrow \quad
x= x_m \equiv 1 / k = (2\pi - \alpha)/(2\pi).
\end{equation}

We assume here and henceforth that $ \widehat{W}\left(x\right)$ is a \emph{strictly convex function}, i.e. $ \widehat{W}''\left(x\right)>0$.
It follows that for any given value of $\lambda_z$, $x= x_m \equiv 1/k$ yields the unique minimum of the strain energy function.
Further, by \eqref{I} we have
\be
I_1(x_m) = 2\lambda_z^{-1} + \lambda_z^2, \qquad
I_2(x_m) = 2\lambda_z + \lambda_z^{-2},
\en
and the minimum value of the strain energy is therefore \emph{independent} of the measure of the opening angle $k \equiv 2\pi/(2\pi-\alpha)$.

From equations \eqref{equil}, \eqref{TT}, and \eqref{change} we have
\be
\dfrac{\text{d}\sigma_{rr}}{\text{d}x} = \dfrac{ \widehat{W}'(x)}{1-x},
\en
and consequently, the radial stress $\sigma_{rr}$ also has a unique extremum, a minimum, at $x=x_m$.
The corresponding relation for the hoop stress $\sigma_{\theta \theta}$ is derivable from \eqref{Tdistrib} and has the form
\be
\dfrac{\text{d}\sigma_{\theta \theta}}{\text{d}x} = \dfrac{ \widehat{W}'(x)}{1-x} + 2 \widehat{W}'(x) + 2 x  \widehat{W}''(x).
\en
In contrast to the radial stress, the determination of its singular point(s) depends on the form of the strain energy function.
Clearly, the hoop stress is increasing at $x = x_m$, the radius where the radial stress is at a minimum.  
At that point, according to \eqref{Tdistrib}$_2$ and \eqref{xm}, $\sigma_{\theta \theta} = \sigma_{rr}$ and, by \eqref{lambdas2}, $\lambda_1 = \lambda_2$; thus the in-plane principal stresses and the stretches are equal there. 
We refer to the surface defined by $x=x_m$ as the \emph{neutral surface}.  
The associated radius ($r$ or $R$) depends, in general, on both $\lambda_z$ and $k$.

Another consequence of \eqref{W'} and convexity is that
\begin{equation} \label{convex}
\left\{
\begin{array}{l}
 \widehat{W}'\left(x\right)<0 \quad \text{for} \quad 0<x<  x_m \equiv 1 /k,
\\[0.1cm]
 \widehat{W}'\left(x\right)>0 \quad \text{for} \quad x > x_m \equiv 1 / k.
 \end{array}
\right.
\end{equation}
Application of these conditions to the tube, however, depends on whether $x_a>x_b$ or $x_b>x_a$.  
It is easy to show from \eqref{gamma}  that either
(i) $x_a>x_b>1$ or (ii) $1>x_b>x_a$.  In case (i) we have $x_m=1/k<1$ so $x_m$ does not lie within the required range and there is no neutral surface in this case.  
In case (ii) we require $x_b>x_m>x_a$.  
Thus, \eqref{convex} can be made more explicit:
\begin{equation} \label{convex1}
\left\{
\begin{array}{l}
 \widehat{W}'\left(x\right)<0 \quad \text{for} \quad x_a<x<  x_m \equiv 1 /k,
\\[0.1cm]
 \widehat{W}'\left(x\right)>0 \quad \text{for} \quad 1>x_b>x > x_m \equiv 1 / k.
 \end{array}
\right.
\end{equation}
This will be used in the next section to determine existence and uniqueness of the unloaded, residually-stressed configuration.

In closing this section, we note that convexity of $ \widehat W(x)$ is a common feature of many standard strain-energy functions, including
the \emph{Mooney-Rivlin model},
\be \label{MR}
 \widetilde W_\text{MR} = \dfrac{C}{2}(I_1-3) + \dfrac{D}{2}(I_2-3), \quad C>0, \quad D \geq 0,
\en
where the material constants $C$ and $D$ have dimensions of stress,
the \emph{Gent model}
\be \label{Gent}
 \widetilde W_\text{G} = -\dfrac{\mu J_m }{2}   \ln\left( 1 - \dfrac{I_1-3}{J_m} \right),  \quad \mu>0, \quad J_m > 0,
\en
and the \emph{Fung model}
\be \label{Fung}
 \widetilde W_\text{F} =\dfrac{\mu}{2c}   \left[\ee^{c(I_1-3)} - 1\right],  \quad \mu > 0, \quad c > 0,
\en
where $\mu$ is the shear modulus in the unstressed configuration and $J_m$ and $c$ are dimensionless constants.  Convexity may easily be checked.
For instance, $ \widehat{W}_\text{MR}''(x) = (C+D\lambda_z^2)/(k \lambda_z x^3)$ which is clearly positive.
Similar, but longer, expressions can be found for $ \widehat{W}_\text{G}''$ and $ \widehat{W}_\text{F}''$, which are also strictly positive.


\section{Existence and uniqueness of the residually-stressed configuration\label{sec3}}


Here and henceforth, we study the residually-stressed tube in configuration $\mathcal{B}_r$.
Setting $P = 0$ in \eqref{P} yields
\begin{equation} \label{P0}
0 = \int _{x_a}^{x_b} \dfrac{ \widehat{W}'(x)}{1- x} \mathrm{d}x,
\quad
x_b = \dfrac{\varepsilon + x_a}{\varepsilon +1}.
\end{equation}
This equation determines an inner radius of the unloaded residually-stressed tube, measured by $x_a$, for a given measure $k$ of the opening angle and given axial stretch $ \lambda_z$.

Two questions need to be answered.
\begin{enumerate}
\item[{(i)}] Does equation \eqref{P0} have a positive solution $x_a$? In other words, is it always possible to close a cylindrical sector in the way indicated when no load is applied to its curved faces?
\item[{(ii)}] If a solution exists, is it unique?
\end{enumerate}

First consider the possible existence of a homogeneous residually-stressed configuration, when the deformation is in the form \eqref{hom}, which is equivalent to assuming
that $x_a = 1$.  Then, by \eqref{Phom}, vanishing of $P$ is equivalent to $ \widehat{W}(1) = 0$. According to \eqref{xm} this leads to $k = 1$, which is not admissible since $k>1$. 
Thus, \emph{there cannot be a homogeneous residually-stressed state of deformation}, i.e.  a residually-stressed configuration is necessarily inhomogeneous.  
This, of course, is well known in general, as shown by, for example, \cite{Hoge85}; see also the discussion in \cite{Ogde03}.

Assume now that $x_a >1$.  Then, as indicated following \eqref{convex}, we must have $x_a > x_b > 1 > x_m$, and hence, according to \eqref{convex}, the integrand in \eqref{P0} has the same sign over the entire range of integration, and there is therefore no solution to \eqref{P0} for $x_a>1$.

For a solution $x_a$ of \eqref{P0} to exist, it must lie in the range $0 < x_a <1$. If either $x_a < x_b < x_m$ or $x_m < x_a < x_b$ then there is again no solution to \eqref{P0}, because the integrand is one-signed over the range of integration. 
Thus, we must have the ordering $x_a < x_m < x_b$ and \eqref{convex1} is applicable. 
Expressing $x_b$ in terms of $x_a$ by means  of \eqref{P0}$_2$ in this latter inequality yields the following range of values for $x_a$:
\begin{equation} \label{range}
[1-(k-1)\varepsilon]x_m < x_a < x_m.
\end{equation}
Then, to ensure that the solution $x_a$ to \eqref{P0} is positive, we enforce the inequality $\varepsilon<1/(k-1)$, i.e.
\be \label{restr}
\dfrac{B^2}{A^2} < \dfrac{k}{k-1} = \dfrac{2\pi}{\alpha},
\en
which prescribes a maximum ratio of outer to inner radius for the original annular sector for a given opening angle, or equivalently a maximum value of the opening angle for a given radius ratio.

Now, addressing question (i) above is equivalent to determining whether the function $f$ defined over the range \eqref{range} by
\be \label{f}
f(y) = \int_{y}^{(\varepsilon+ y)/(\varepsilon+1)}
\dfrac{ \widehat{W}'(x)}{1-x} \mathrm{d}x,\quad [1-(k-1)\varepsilon]x_m < y < x_m,
\en
has a zero. First we examine the value of $f(y)$ at the upper bound of the range, which is
\be \label{f1}
f(x_m) = \int_{x_m}^{(\varepsilon+x_m)/(\varepsilon+1)}
\dfrac{ \widehat{W}'(x)}{1- x} \mathrm{d}x ,
\en
and this is positive because
\be
x_m < \dfrac{\varepsilon+x_m}{\varepsilon+1} < 1,
\en
so that the integrand in \eqref{f1} is always positive over the range of integration. 
Next, we examine $f(y)$  at the lower bound of the range \eqref{range};
after some simplification, we find that
\be \label{f2}
f\left([1-(k-1)\varepsilon]x_m\right) = \int_{[1-(k-1)\varepsilon]x_m}^{x_m}
\dfrac{ \widehat{W}'(x)}{1-x} \mathrm{d}x,
\en
which is strictly negative because
\be
[1-(k-1)\varepsilon]x_m < x_m < 1,
\en
so that the integrand in \eqref{f1} is negative over the whole range of integration.
Because $f$ is continuous we conclude that there exists at least one solution $y=x_a$ of the equation $f(y) = 0$ in the range \eqref{range}.

The answer to question (ii) can be found by computing $f'(y)$, which yields
\be
f'(y) = \dfrac{\widehat{W}'\left( (\varepsilon+ y)/(\varepsilon+1)\right) - \widehat{W}'(y)}{1- y}.
\en
Because $\widehat{W}'$ is a monotonic increasing function, and because of the ordering
\be
y < \dfrac{\varepsilon+ y}{\varepsilon+1} < 1,
\en
it follows that $f'(y)>0$ for all $y$ in the range \eqref{range}. Thus the answer to question (ii) is also positive.

In conclusion, an annular sector can always be closed into a tube with an accompanying axial stretch if its strain-energy function $\widetilde{W}(x)$ is strictly convex and its thickness is small enough, as prescribed by \eqref{restr}.


\subsection{Thin-walled tubes}


We have shown in the previous section that if the strain-energy function is strictly convex, then a unique residually-stressed configuration exists for all such energy functions and for any given opening angle and axial stretch. 
The dimensions of this residually-stressed tube depend on the form of the strain-energy function, the initial geometry of the unstressed sector, and the value of the axial stretch. 
For thin tubes, some general statements can be made about the form of the residually-stressed configuration, as follows.

Expanding $f(y)$ in \eqref{f}  as a Maclaurin series in the thickness parameter $\varepsilon$, substituting into the equation $f(x_a)=0$, and retaining only the leading-order term, yields the equation
\be
\widehat{W}'\left(x_a\right) = 0,
\en
which, by \eqref{xm}, has the unique solution $x_a=x_m=1/k$. 
Thus, thin tubes have the nice property that the residually-stressed configuration $\mathcal{B}_r$ corresponds to the minimum of the strain energy. 

Moving on to thicker tubes, we expand equation  \eqref{P0} to the next order in $\varepsilon$ and this yields
\be\label{thicker}
 \widehat{W}'(x_a) +\dfrac{1}{2}\varepsilon\left[(1-x_a)\widehat{W}''(x_a)-\widehat{W}'(x_a)\right]=0.
\en
Then we look for $x_a$ in the neighborhood of $x_m$, setting $x_a = x_m\left[1+\varepsilon \delta + \mathcal{O}(\varepsilon^2)\right]$, say.
Substitution of this into the equation above yields
\be
\left[x_m \delta + \dfrac{1}{2}\left(1-\dfrac{1}{k}\right)\right]\widehat{W}''(x_m)=0,
\en
but $\widehat{W}''>0$, so that $x_m \delta = -(k-1)/(2k)$, and hence
\be
x_a = x_m\left[1 - \varepsilon(k-1)/2 + \mathcal{O}(\varepsilon^2)\right],
\en
showing that a thicker initial sector leads to a smaller inner radius of the residually-stressed tube (a result independent of the particular form of strain-energy function within the considered class).


\subsection{Examples}
\label{Examples}


We now return to the case of tubes with finite thickness. 
In the preceding analysis, $\varepsilon \equiv (B/A)^2-1$ was used as the thickness parameter.
It gives a measure of the thickness of the unloaded, unstressed annular sector $\mathcal{B}_0$.
It follows easily from simple kinematics that the thickness ratio of the residually-stressed tube in configuration $\mathcal{B}_r$, $\hat\varepsilon$ say, is related to   $\varepsilon$ by
\be \label{gamma0}
\hat\varepsilon \equiv \left(\dfrac{b}{a}\right)^2 -1 = \dfrac{\varepsilon}{x_a} > \varepsilon.
\en
To make further progress, the strain-energy function has to be specified.

\begin{figure}[!t]
\center
\epsfig{figure=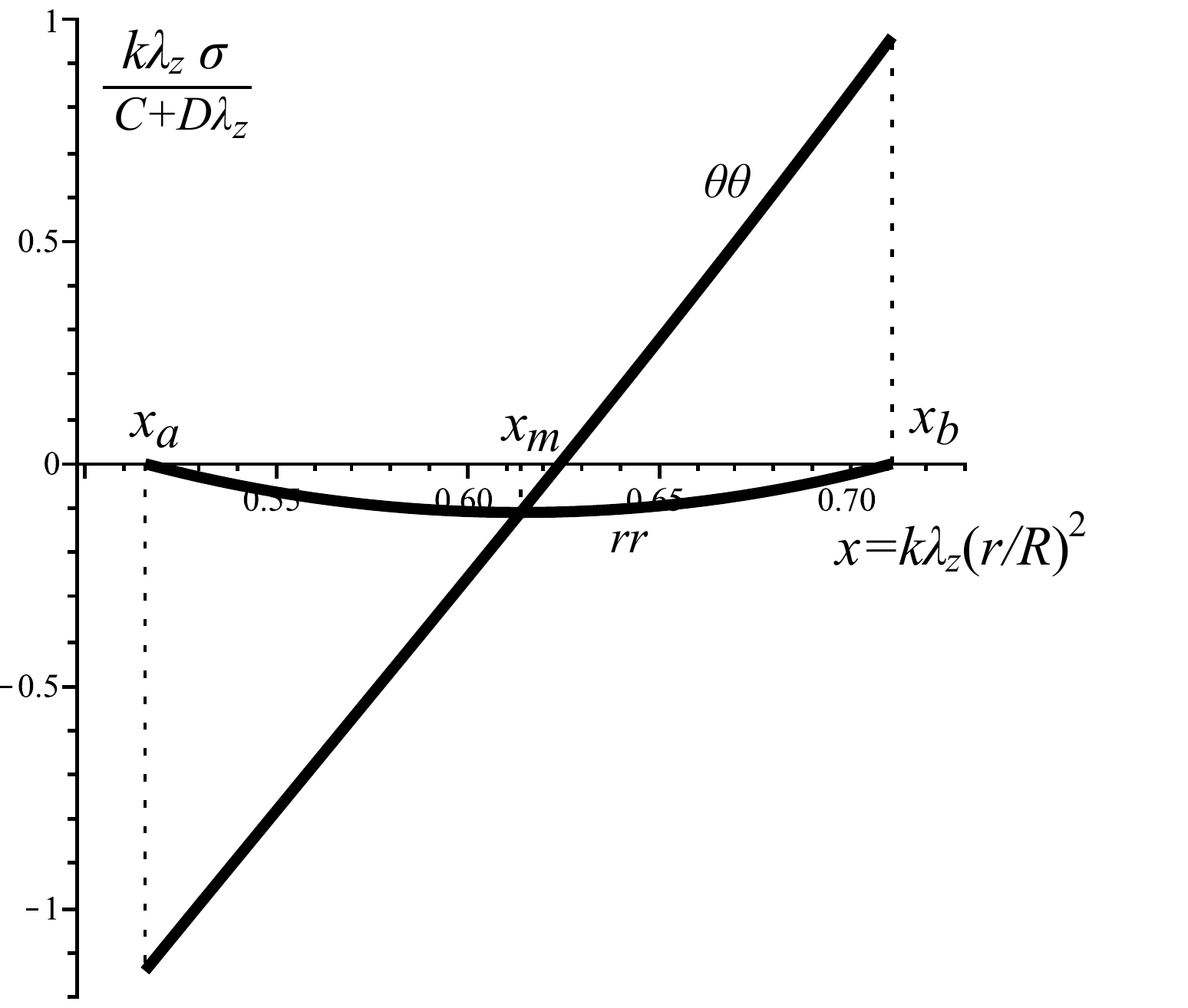, width=.7\textwidth}
 \caption{Variation of the non-dimensional radial stress $\sigma_{rr}$ and hoop stress $\sigma_{\theta \theta}$ through an arterial wall modeled as a Mooney-Rivlin material. 
 The tube is in a residually-stressed configuration.  
 The opening angle is taken as $\alpha=139^\circ$.
 }
 \label{figure2}
\end{figure}

First, we consider the Mooney-Rivlin model \eqref{MR}.
For this material, it is found that the defining equation \eqref{P0} for the residually-stressed state  is independent of the material constants $C$ and $D$ and does not depend explicitly on the axial stretch $\lambda_z$; it is given by
\be \label{solve1}
\ln \left(\dfrac{\varepsilon + x_a}{x_a}\right) -k^2 \ln\left(\varepsilon+1\right)
+ \dfrac{\varepsilon(1-x_a)}{x_a(\varepsilon + x_a)} = 0,
\en
or, equivalently, in terms of  $\hat\varepsilon$,
\be \label{solve2}
\ln \left(\hat\varepsilon + 1\right)  -  k^2 \ln\left(1 + \hat\varepsilon x_a\right) + \dfrac{\hat\varepsilon}{\hat\varepsilon+1}\left(\dfrac{1-x_a}{x_a} \right) = 0.
\en
The constants in these equations now need to be specified.
\cite{MaSa02} measured the inner and outer radii of five tubular segments of bovine thoracic aortas, with mean values of $a= 11.48$ mm and $b = 17.45$ mm, respectively, yielding $\hat \varepsilon = 1.310$.
On cutting these segments longitudinally, they  found an average opening angle of $\alpha = 139^\circ$, which yields $k = 1.629$.
Substituting these values into \eqref{solve2} and solving the equation numerically yields $x_a= 0.5157$.
This value and \eqref{gamma0}$_2$ then yield $\varepsilon = 0.6759$ which, incidentally, is indeed smaller than $1/(k-1) = 1.590$, in line with the requirement \eqref{restr}.
Finally, the outer value of $x$ is determined as $x_b = 0.7110$ by \eqref{P0}$_2$, and the radial stress reaches its minimum at $x=x_m = 0.6139$.
In Figure \ref{figure2}, we plot the variations of the resulting (residual) stresses $\sigma_{rr}$ and $\sigma_{\theta \theta}$, divided by the quantity $(C+D\lambda_z^2)/(k\lambda_z)$, as a function of $x$.
These figures are entirely consistent with those produced by other authors, as in, for example, \cite{Ogde03}, \cite{Ragh04} and \cite{Olss06}.

\color{black}
For our second example, we use data from an article by \cite{Delf97}, who consider measurements made on human carotid arteries.
In their experimental, theoretical, and numerical investigation of the mechanical behavior of arteries, \cite{Delf97} model the arterial wall as a homogeneous, hyperelastic, isotropic solid, exhibiting residual stress and strain-stiffening effects.
They use the opening angle method to account for the former and the Fung model for the latter.
\color{black}
They report the following typical values: $a = 3.1$ mm, $b=4.0$ mm, and $\alpha = 100^\circ$.
These give here: $\hat \varepsilon = 0.6649$ and $k = 1.385$.
\cite{Delf97} go on to fit pressure measurements to the predictions of the Fung strain energy density \eqref{Fung} and find the following material parameters:
$\mu = 44.2$ kPa, $c=8.35$.
It follows that $x_a$ can be determined by solving \eqref{P0}, which reads here as
\begin{equation}
0 = \int _{x_a}^{x_b} \dfrac{ \widehat{W}_F'(x)}{1- x} \mathrm{d}x,
\quad
x_b = \dfrac{(\hat\varepsilon +1) x_a}{\hat \varepsilon x_a +1}.
\end{equation}
Numerically, we find that $x_a = 0.6701$.
We may then deduce the expected value for the radius of the opened sector to be $A= a \sqrt{k \lambda_z/x_a} = 4.673$ mm.
This comes within less than 5\% of the experimentally measured radius, which is $4.41$ mm \citep{Delf97}.


\section{Stability of the residually-stressed configuration\label{sec4}}


In the absence of body forces the equilibrium equation can be written quite generally as
\begin{equation}
\Div\vec{S}=\vec{0},\label{S-equil}
\end{equation}
where $\vec{S}$ is the nominal stress tensor and $\Div$ is the divergence operator with respect to $\mathcal{B}_0$.  
For an elastic material with strain-energy function $W(\vec{F})$ and subject to the incompressibility constraint $\det\vec{F}\equiv 1$, the nominal stress is given by
\begin{equation}
\vec{S}=\frac{\partial W}{\partial\vec{F}}-q\vec{F}^{-1},\label{S-const}
\end{equation}
where $q$ is again the Lagrange multiplier associated with the constraint.  
In terms of the Cauchy stress tensor $\vec{\sigma}$ the equilibrium equation has the equivalent form $\div\vec{\sigma}=\vec{0}$ and we note the connection $\vec{\sigma}=\vec{FS}$.  
It was the specialization of the latter form of the equilibrium equation to the cylindrical geometry that was used in Section 2.2.  In considering the stability of the configuration $\mathcal{B}_r$ it is convenient to work in terms of the incremental form of the equilibrium equation \eqref{S-equil} and the incremental form of the constitutive equation \eqref{S-const}, which we now summarize.


\subsection{Incremental equations}


Let $\vec{u}$ be a small displacement from the configuration $\mathcal{B}_r$ and let $\vec{\dot{S}}$ be the associated increment in $\vec{S}$.  
The (linearized) incremental form of the constitutive equation \eqref{S-const} is written as
\begin{equation}
\vec{\dot{S}}=\vec{\mathcal{A}}\vec{\dot{F}}+q\vec{\dot{F}}-\dot{q}\vec{F}^{-1},
\end{equation}
where $\vec{\dot{F}}$ and $\dot{q}$ are the increments in $\vec{F}$ and $q$, respectively, and $\vec{\mathcal{A}}$ is the fourth-order tensor of fixed-reference elastic moduli, defined by
\begin{equation}
\vec{\mathcal{A}}=\frac{\partial^2W}{\partial\vec{F}\partial\vec{F}},\quad \mathcal{A}_{\alpha i\beta j}=\frac{\partial^2W}{\partial F_{i\alpha}\partial F_{j\beta}}.
\end{equation}
The incremental form of the equilibrium equation is then
\begin{equation}
\Div\vec{\dot{S}}=\vec{0}.
\end{equation}

Let $\vec{x}$ denote the position vector of a point in $\mathcal{B}_r$.  
Then, by updating the reference configuration from $\mathcal{B}_0$ to $\mathcal{B}_r$, we may write the incremental equilibrium equation in Eulerian form as
\begin{equation}
\div\vec{\dot{S}}_0=\vec{0},\label{increm0equil}
\end{equation}
where
\begin{equation}
\vec{\dot{S}}_0=\vec{\mathcal{A}}_0\vec{\Gamma}+q\vec{\Gamma}-\dot{q}\vec{I},\label{dotS0}
\end{equation}
is the push forward $\vec{F{\dot{S}}}$ of $\vec{\dot{S}}$, $\vec{\Gamma}$ is the displacement gradient $\grad\vec{u}$, $\vec{I}$ is the identity tensor, and $\vec{\mathcal{A}}_0$, the push-forward of $\vec{\mathcal{A}}$, is the tensor of instantaneous moduli, whose components are given by $\mathcal{A}_{0piqj}=F_{p\alpha}F_{q\beta}\mathcal{A}_{\alpha i\beta j}$.
For full details of these connections, see \citep{Ogde84}.  The incremental incompressibility condition is then
\begin{equation}
\tr{\vec{\Gamma}}\equiv\div\vec{u}=0.
\end{equation}

In the configuration $\mathcal{B}_r$ it is appropriate to work in terms of cylindrical polar coordinates $r,\theta,z$ and their associated basis vectors $\vec{e}_r,\vec{e}_\theta,\vec{e}_z$.  
Let $(u,v,w)$ be the components of $\vec{u}$.  
Then the matrix of components of $\vec{\Gamma}$ with respect to this basis is
\begin{equation}
\begin{bmatrix}
 u_{,r} & (u_{,\theta} - v)/r & u_{,z} \\
 v_{,r} & (u + v_{,\theta})/r & v_{,z}  \\
 w_{,r} & w_{,\theta}/r       & w_{,z}   \end{bmatrix}\label{Gammamatrix}
\end{equation}
and the incremental incompressibility condition becomes
\begin{equation}
u_{,r}+\frac{u+v_{,\theta}}{r}+w_{,z}=0.\label{inc3d}
\end{equation}

Referred to principal axes of $\vec{B}$ the non-zero components of $\vec{\mathcal{A}}_0$ are
\citep{Ogde84}:
\begin{align} \label{A0_principal}
& \mathcal{A}_{0iijj} =
\lambda_i \lambda_j W_{ij},
\nonumber \\
& \mathcal{A}_{0ijij} =
  (\lambda_i W_i-\lambda_j
W_j)\lambda_i^2/(\lambda_i^2-\lambda_j^2),\quad i\neq j,
\nonumber \\
& \mathcal{A}_{0ijji}= \mathcal{A}_{0jiij}=
   \mathcal{A}_{0ijij} - \lambda_i W_i,\quad i\neq j,
\end{align}
(no sums on repeated indexes here),
where $W_j \equiv \partial W / \partial \lambda_j$ and
$W_{i j} \equiv \partial^2 W / \partial \lambda_i \partial
\lambda_j$.

In the present situation the cylindrical axes are principal axes of $\vec{B}$ and the above apply with $i$ and $j$ running over values $r$, $\theta$, and $z$.  
The components $\dot{S}_{0ij}$ are then easily obtained from \eqref{dotS0} and \eqref{Gammamatrix}.  
Explicit expressions for the incremental equations of equilibrium for a thick-walled tube in cylindrical polar coordinates can be found in \cite{Haug79}.  
Here, however, for purposes of illustration and for simplicity we now specialize the incremental equations so that $w=0$ and $u$ and $v$ are independent of $z$, so that we consider prismatic incremental deformations.  
The equilibrium equation \eqref{increm0equil} reduces to two component equations, which can be written as
\begin{equation}
(r\dot{S}_{0rr})_{,r}+\dot{S}_{0\theta r,\theta}-\dot{S}_{0\theta\theta}=0, \quad
(r\dot{S}_{0r\theta})_{,r}+\dot{S}_{0\theta \theta,\theta}+\dot{S}_{0\theta r}=0,
\end{equation}
and from \eqref{dotS0} we obtain
\begin{eqnarray}
\dot{S}_{0rr}&=&(\mathcal{A}_{0rrrr}+q)u_{,r}+\mathcal{A}_{0rr\theta\theta}\dfrac{u+v_{,\theta}}{r}-\dot{q},\notag \\
\dot{S}_{0r\theta}&=&\mathcal{A}_{0r\theta r\theta}v_{,r}+(\mathcal{A}_{0r\theta r\theta}-\sigma_{rr})\dfrac{u_{,\theta}-v}{r},\notag \\
\dot{S}_{0\theta r}&=&\mathcal{A}_{0\theta r\theta r}\dfrac{u_{,\theta}-v}{r}+(\mathcal{A}_{0\theta r\theta r}-\sigma_{\theta\theta})v_{,r},\notag \\
\dot{S}_{0\theta\theta}&=&(\mathcal{A}_{0\theta\theta\theta\theta}+q)\dfrac{u+v_{,\theta}}{r}+\mathcal{A}_{0rr\theta\theta}u_{,r}-\dot{q},
\end{eqnarray}
and the incremental incompressibility equation \eqref{inc3d} reduces to
\begin{equation}
u_{,r}+\frac{u+v_{,\theta}}{r}=0.\label{inc2d}
\end{equation}
Note that use has been made of the connections
\begin{equation}
\mathcal{A}_{0r\theta \theta r}+q=\mathcal{A}_{0r\theta r\theta}-\sigma_{rr}=\mathcal{A}_{0\theta r\theta r}-\sigma_{\theta\theta},
\end{equation}
for which equations \eqref{T} and \eqref{A0_principal} have been utilized.

Note that the components of incremental traction on a surface $r =\mbox{constant}$ are $\dot{S}_{0rr}$ and $\dot{S}_{0r\theta}$.


\subsection{Stroh formulation}


We now consider solutions of the form
\begin{equation}
\{u,v,\dot{q}\}=\{U(r),V(r),Q(r)\}\mathrm{e}^{\mathrm{i}n\theta},
\end{equation}
where $U$, $V$, and $Q$ are functions of $r$ only, so that the solutions are single valued, and $n=0,1,2,3, \ldots$ is the circumferential number.  
It follows that the components $\dot{S}_{0ij}$ have a similar structure and we therefore write
\begin{equation}
\dot{S}_{0ij}=\Sigma_{ij}(r)\mathrm{e}^{\mathrm{i}n\theta},
\end{equation}
say, where $\Sigma_{ij},\, i,j\in\{r,\theta\}$ are functions of $r$ only.
Then we obtain
\begin{eqnarray}
\Sigma_{rr}&=&(\mathcal{A}_{0rrrr}+p)U'+\frac{1}{r}\mathcal{A}_{0rr\theta\theta}(U+\mathrm{i}nV)-Q, \notag \\
\Sigma_{r\theta}&=&\mathcal{A}_{r\theta r\theta}V'+\frac{1}{r}(\mathcal{A}_{r\theta r\theta}-\sigma_{rr})(\mathrm{i}nU-V),\label{secondSigma}\notag \\
\Sigma_{\theta r}&=&\frac{1}{r}\mathcal{A}_{0\theta r\theta r}(\mathrm{i}nU-V)+(\mathcal{A}_{0\theta r\theta r}-\sigma_{\theta\theta})V',\notag \\
\Sigma_{\theta\theta}&=&\frac{1}{r}(\mathcal{A}_{0\theta\theta\theta\theta}+p)(U+\mathrm{i}nV)+\mathcal{A}_{0rr\theta\theta}U'-Q,
\end{eqnarray}
and the incremental incompressibility condition \eqref{inc2d} yields
\begin{equation}
U'=-\frac{1}{r}U-\frac{\mathrm{i}n}{r}V.
\end{equation}
From \eqref{secondSigma} we obtain
\begin{equation}
V'=-\frac{\mathrm{i}n}{r}(1-\sigma)U+\frac{1}{r}(1-\sigma)V+\frac{\Sigma_{r\theta}}{\alpha},
\end{equation}
where we have introduced the notations
\begin{equation} \label{sigma}
 \alpha =\mathcal{A}_{0r\theta r\theta}, \qquad
 \sigma=\sigma_{rr}/\alpha.
\end{equation}
By using the above and then eliminating $\Sigma_{\theta r}$, $\Sigma_{\theta\theta}$, and $Q$ from the equilibrium equations in favor of $U$, $V$, $\Sigma_{rr}$, and $\Sigma_{r\theta}$, the two equilibrium equations yield expressions for $(r\Sigma_{rr})'$ and $(r\Sigma_{r\theta})'$ in terms of $U$, $V$, $\Sigma_{rr}$, and $\Sigma_{r\theta}$.  
Then, by introducing the four-component displacement-traction vector $\vec{\eta}$ defined by
\begin{equation}
\vec{\eta}=[U,V,\mathrm{i}r\Sigma_{rr},\mathrm{i}r\Sigma_{r\theta}]^T,
\end{equation}
we may cast the governing equations in the Stroh form \citep{Shuv03}
\be \la{stroh}
\dfrac{\text{d}}{\text{d} r} \vec{\eta}(r) = \dfrac{\ii}{r} \vec{G}(r) \vec{\eta}(r),
\en
where the matrix $\vec{G}$ has the Stroh structure,
\be \label{G}
\vec{G} = \begin{bmatrix} 
  \ii & -n & 0 & 0 \\
  -n (1 - \sigma) &  - \ii (1 - \sigma) & 0  &  -  1/\alpha\\
  \kappa_{11} & \ii \kappa_{12} & -\ii & -n (1 - \sigma) \\
  - \ii \kappa_{12} & \kappa_{22}& -n &  \ii (1 - \sigma)  
          \end{bmatrix}
\en
where
\begin{align}
& \kappa_{11} = 2 \beta + 2\alpha (1- \sigma) 
  + n^2\left[\gamma - \alpha(1 - \sigma)^2\right],
 \notag \\[2pt]
& \kappa_{12} = n \left[2 \beta + \gamma + \alpha(1- \sigma^2)\right],
 \notag \\[2pt]
& \kappa_{22} =  
   \gamma - \alpha(1 - \sigma)^2
   + 2 n^2[\beta + \alpha(1-\sigma)],
\end{align}
and the notations $\gamma$ and $\beta$ are defined by
\begin{equation}
\gamma=\mathcal{A}_{0\theta r\theta r},\quad 2\beta=\mathcal{A}_{0rrrr}+\mathcal{A}_{0\theta\theta\theta\theta}-2\mathcal{A}_{0rr\theta\theta}-2\mathcal{A}_{0r\theta\theta r}.
\end{equation}
Note that $\kappa_{11}+\kappa_{22}=(n^2+1)\kappa_{12}/n$.
Note also that this form of the Stroh formulation is entirely consistent with previous forms obtained in the bending of blocks \citep{DeNC09,  DGMM10}, or the compression of tubes \citep{GoVD08}, the only differences being in the actual expressions of the moduli.

The above coefficients may also be conveniently expressed in terms of $\widehat{W}(x)$ through the connections
\begin{equation}
\alpha=\frac{2x\widehat{W}'(x)}{k^2 x^2 - 1}, \qquad 
\gamma = k^2 x^2 \alpha,
\qquad 
\beta = 2x^2\widehat{W}''(x)+x\widehat{W}'(x) - \alpha, 
\end{equation}
and $\sigma$ is given by \eqref{Tdistrib}$_1$ and \eqref{sigma}.

We may now devise a numerical strategy to find out whether a given open sector remains stable once closed into a tube, as follows.
The prescribed geometrical quantities are the measure of the initial sector thickness $\varepsilon$, the measure of the opening angle $k$, and the axial stretch $\lambda_z$.
The prescribed physical quantity is the strain energy density  $\widehat{W}(x)$.
These quantities are required to determine $x_a$ and $x_b$ from \eqref{P0}.
Then we integrate numerically the incremental equations \eqref{stroh}, written as
\be \la{stroh-x}
\dfrac{\text{d}}{\text{d} x} \vec{\eta}(x) = \dfrac{\ii}{2x(1-x)} \vec{G}(x) \vec{\eta}(x),
\en
(using \eqref{change}), and we check whether (stability) or not (instability) the inner and outer faces of the tube are left free of incremental traction, i.e. whether or not
\be
\vec{\eta}(x_a) = [U(x_a), V(x_a), 0, 0]^T, \qquad
\vec{\eta}(x_b) = [U(x_b), V(x_b), 0, 0]^T. 
\en
This two-point boundary-value problem may run into numerical stiffness problems, especially for thick sectors, and we use the compound matrix method \citep{HaOr97, DeNC09} and the impedance matrix method \citep{DeNC09} to take care of these.


\subsection{Example: Mooney-Rivlin tubes}


Motivated by simple experimental observations on sectors made of silicone rubbers, see Figure \ref{figure2}, we now focus on tubes made of solids with the Mooney-Rivlin strain-energy density \eqref{MR}.
We find that 
\be
\alpha = (C\lambda_z^{-1} + D\lambda_z)\dfrac{1}{k x}, \qquad
\gamma =  (C\lambda_z^{-1} + D\lambda_z)k x, \qquad
2 \beta = \alpha + \gamma,
\en
and
\be
\sigma = \dfrac{x}{2}\left[(1-k^2)\ln \left(\dfrac{x-1}{x_b-1}\right) - \ln \left(\dfrac{x}{x_b}\right) + \dfrac{1}{x} - \dfrac{1}{x_b}\right].
\en
In the latter equation, $x_b$ is determined by solving \eqref{solve1} for $x_a$, a calculation which is done independently of $C$, $D$, and $\lambda_z$.

\begin{figure}[!t]
\center
\epsfig{figure=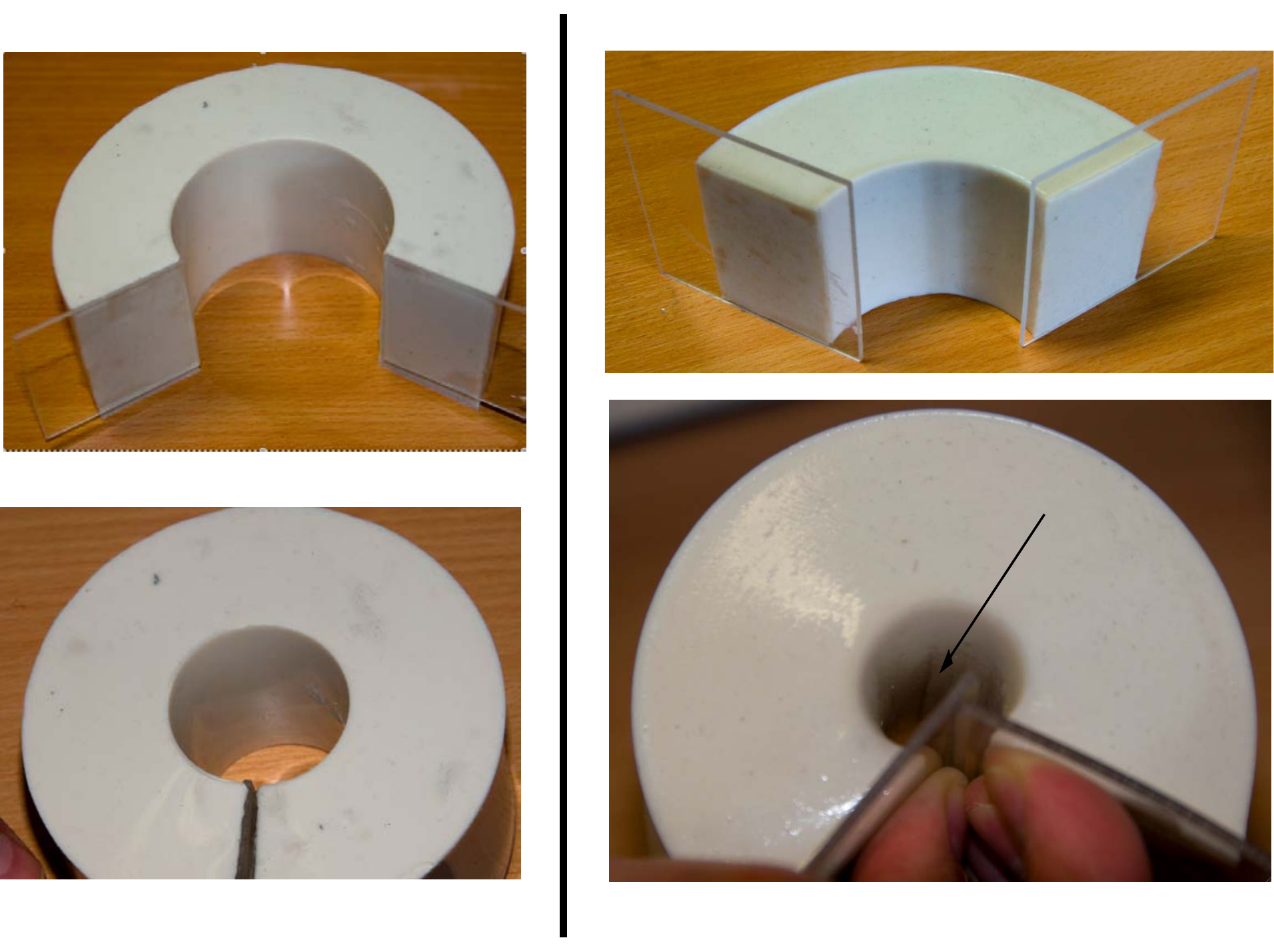, width=.9\textwidth}
 \caption{Annular sectors made with rubber silicone; Height: 6 cm; Inner radius: $A = 3.5$ cm; Outer radius: $B=7.2$ cm. 
 Two rectangles of acrylic glass are glued to the end faces of the sectors in order to bend them.
 On the left: an  unstressed open sector with opening angle $\alpha = \pi/3$ is closed into a residually stressed and unstretched tube, and its inner curved face remains free of wrinkles. 
 On the right: an open sector with opening angle $\alpha = 2\pi/3$ is bent but cannot be closed without the appearance of axial wrinkles on its inner face, one of which is marked by an arrow. }
 \label{figure3}
\end{figure}

It is now clear upon inspection of the incremental equations \eqref{stroh-x} that once we normalize the third and fourth components of $\vec{\eta}$ with respect to the quantity $(C\lambda_z^{-1} + D\lambda_z)$, we end up with a completely non-dimensional system of differential equations which is independent of the material parameters $C$ and $D$ and of the axial stretch $\lambda_z$.
Only three  parameters are then required to determine the stability of closed-up Mooney-Rivlin tubes, namely the following geometrical quantities:
$B/A$, the ratio of the initial outer to inner radii; $\alpha$, the opening angle; and $n$, the mode number.

\begin{figure}[!t]
\center
\epsfig{figure=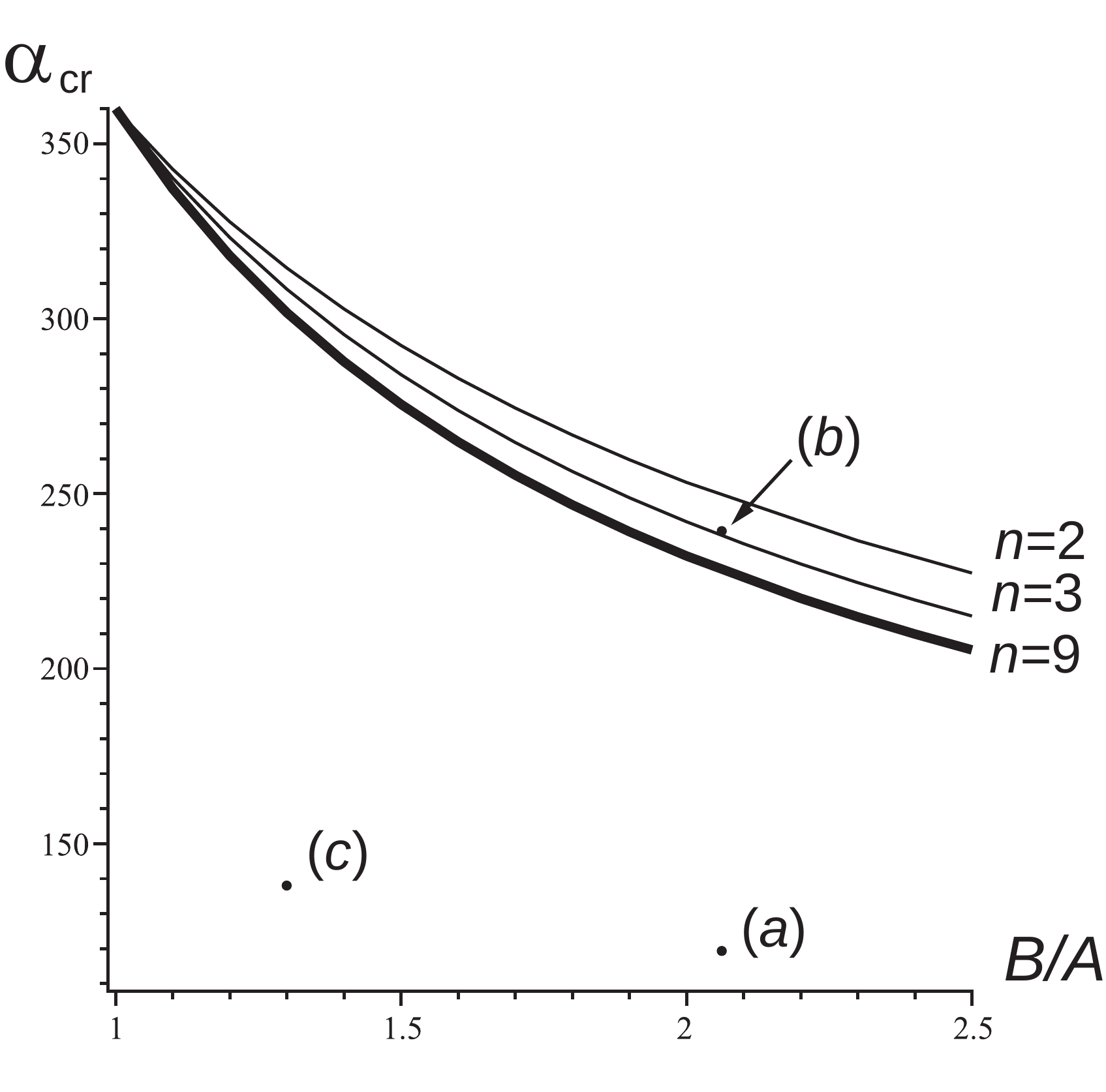, width=.8\textwidth}
 \caption{Instability of a closed tube with Mooney-Rivlin strain energy: critical opening angle $\alpha_\text{cr}$ versus initial radii ratio $B/A$, for different mode numbers $n$.
The lowest curve --the bifurcation curve-- corresponds to modes $n=8$, $9$, $10$ (almost indistinguishable one from another).
The curves of all other modes are situated above this ($n=2$, $3$ are shown).
The annular silicone sector on the left in Figure \ref{figure3} and the opened artery of Section \ref{Examples} correspond to stable geometries (Points  \emph{(a)} and \emph{(c)}). 
The annular silicone sector on the right in Figure \ref{figure3} gives a point above the bifurcation curve, Point \emph{(b)}.}
 \label{figure4}
\end{figure}

Figure \ref{figure4} shows the variations of the critical opening angle $\alpha_\text{cr}$ (in degrees) with $B/A$, for different values of $n$.
The lowest curve represents the \emph{bifurcation curve}, below which all tubes are stable.
In our numerical investigation, we found that this corresponds to the $n=9$ curve, although the $n=8$ and $n=10$ curves are almost indistinguishable from it. 
It follows that when a closed-up Mooney-Rivlin tube buckles, it presents at least 8 wrinkles on its inner curved face.
When the initial sector is thin, $B/A \rightarrow 0$, it can be bent into a tube without buckling, so that $\alpha_\text{cr} \rightarrow 360^\circ$ in that limit, for all mode numbers.
This is consistent with the observation that thin rectangular plates with a neo-Hookean strain energy can be bent into full circles without encountering bifurcation \citep{Haug99}.

The radii ratio of the silicone sectors in Figure \ref{figure3} is $B/A = 2.06$.
We see that an opening angle of $\alpha = 120^\circ$ is below the bifurcation curve, so that the corresponding sector can be closed into a tube in a stable manner; see Point \emph{(a)} in Figure \ref{figure4}.
For an opening angle of $\alpha = 240^\circ$, we are above the bifurcation curve (see Point \emph{(b)} in Figure \ref{figure4}), and wrinkles are thus predicted. 
Finally, the data of \cite{MaSa02}, fitted to the Mooney-Rivlin strain energy, give $B/A = \sqrt{1 + x_a \hat{\varepsilon}} = 1.29$; see Section \ref{Examples}. 
In that case, their measured opening angle of $\alpha = 139^\circ$ is clearly below the bifurcation curve, see Point \emph{(c)} in Figure \ref{figure4}, and the artery can be ``closed'' in a stable manner.

\color{black}

\section{Conclusion}


We have proved that the opening angle method gives a unique solution to the equations of equilibrium provided the strain energy density of the solid is a convex function of the geometrical quantity $x = 2\pi \lambda_z(r^2/R^2)/(2\pi-\alpha)$, where $\lambda_z$ is the axial pre-stretch, $\alpha$ is the opening angle, and $r$, $R$ are the initial and current radial coordinates, respectively.
The convexity is indeed met by many standard strain energy densities including the Mooney-Rivlin, Fung, and Gent models.
We have also provided an analysis of incremental prismatic deformations superimposed on the finite deformation in order to assess the stability of the intact tube formed by the closing of the sector.

The results are valid for homogeneous isotropic solids and are thus highly relevant to the mechanical modeling of the closing of annular sectors into tubes made of rubber-like materials. 
This correlation was confirmed by our toy experiments with sectors made of soft silicone. 
With respect to biomechanics, where the opening angle method is a tool widely used for the estimation of residual stresses in arteries, we note that there are many articles in the literature where arterial walls are modeled as being homogeneous and isotropic, as, for example, in  \cite{Halp91} or \cite{Delf97}.
However, it is now well established and accepted that arteries are inhomogeneous and anisotropic solids, due, in particular, to a layered structure and the presence of collagen fiber bundles. 
The impact of such inhomogeneity and anisotropy on our results remains to be assessed.

\color{black}



\section*{Acknowledgements}


This work is supported by a Senior Marie Curie Fellowship awarded by the Seventh Framework Programme of the European Commission to the first author, and by an E.T.S. Walton Award given to the third author by Science Foundation Ireland.
This material is based upon works supported by the Science Foundation Ireland under Grant No. SFI 08/W.1/B2580.

We are most grateful to Stephen Kiernan for taking the pictures of Figure~\ref{figure3}.




\bibliographystyle{elsarticle-harv}

\end{document}